**Atomic Gold Ions Clustered with Noble Gases: Helium, Neon, Argon, Krypton, and Xenon**

Paul Martini,[1] Lorenz Kranabetter,[1] Marcelo Goulart,[1] Bilal Rasul,[1,2] Michael Gatchell,[1,3] Paul Scheier,[1*] Olof Echt[1,4*]

[1] Institut für Ionenphysik und Angewandte Physik, Universität Innsbruck, Technikerstr. 25, A-6020 Innsbruck, Austria
[2] Department of Physics, University of Sargodha, 40100 Sargodha, Pakistan
[3] Department of Physics, Stockholm University, 106 91 Stockholm, Sweden
[4] Department of Physics, University of New Hampshire, Durham NH 03824, USA

\* Corresponding authors:
Paul Scheier <paul.scheier@uibk.ac.at>
Olof Echt <olof.echt@unh.edu>

ORCID IDs:
Olof Echt 0000-0002-0970-1191
Michael Gatchell 0000-0003-1028-7976
Marcelo Goulart 0000-0002-6006-9339
Paul Scheier 0000-0002-7480-6205

Abstract
High-resolution mass spectra of helium droplets doped with gold and ionized by electrons reveal $He_nAu^+$ cluster ions. Additional doping with heavy noble gases results in $Ne_nAu^+$, $Ar_nAu^+$, $Kr_nAu^+$, and $Xe_nAu^+$ cluster ions. The high stability predicted for covalently bonded $Ar_2Au^+$, $Kr_2Au^+$, and $Xe_2Au^+$ is reflected in their relatively high abundance. Surprisingly, the abundance of $Ne_2Au^+$ which is predicted to have zero covalent bonding character and no enhanced stability features a local maximum, too. The predicted size and structure of complete solvation shells surrounding ions with essentially non-directional bonding depends primarily on the ratio $\sigma^*$ of the ion-ligand *versus* the ligand-ligand distance. For $Au^+$ solvated in helium and neon the ratio $\sigma^*$ is slightly below 1, favoring icosahedral packing in agreement with a maximum observed in the corresponding abundance distributions at $n = 12$. $He_nAu^+$ appears to adopt two additional solvation shells of $I_h$ symmetry, containing 20 and 12 atoms, respectively. For $Ar_nAu^+$, with $\sigma^* \approx 0.67$, one would expect a solvation shell of octahedral symmetry, in agreement with an enhanced ion abundance at $n = 6$. Another anomaly in the ion abundance at $Ar_9Au^+$ matches a local maximum in its computed dissociation energy.

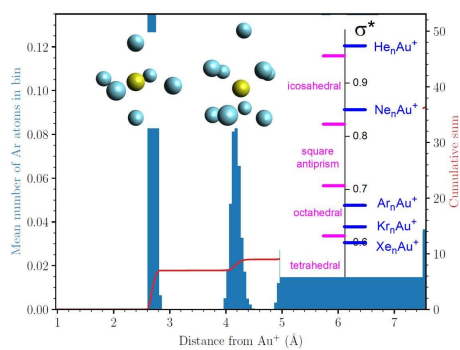

Graphic for Table of Content

## 1. Introduction

In 1995 Pyykkö predicted that $Au^+$ forms chemical bonds with the heavy noble gases (Ng) argon, krypton, and xenon.[1] For $XeAu^+$, the most strongly bound diatomic, Pyykkö calculated a bond energy $D_e = 0.91$ eV, over half of it coming from relativistic effects. For $[XeAuXe]^+$ he computed an atomization energy of 2.25 eV and a bond length approaching the sum of the covalent radii of Xe and Au, respectively. Read and Buckingham questioned the need to invoke covalent bonding, arguing that inclusion of dispersion and higher order polarization effects will account for strong bonding.[2]

The nature of the bond in neutral and cationic $NgAu^{0,+}$ is still being debated;[3-5] it has even been argued that it is a distinct kind of bond.[6] The strength of the bond in $XeAu^+$, however, is beyond doubt. Schröder et al. published coupled cluster calculations of $XeAu^+$ with single and double excitations and a perturbative treatment of the triples (CCSD(T)) with full counterpoise corrections that resulted in a dissociation energy $D_e = 1.31$ eV.[7]

The bond energy and bond length of neutral and charged $NgAu^{\pm,0}$ (as well as $NgAg^{\pm,0}$ and $NgCu^{\pm,0}$) have been the subject of numerous other high-level theoretical studies;[8-17] also see the recent review by Pan et al..[18] One interesting, widely shared conclusion is that physical interaction is sufficient to explain bonding in the weakly bound $HeAu^+$ and $NeAu^+$ ($D_e = 0.048$ eV and 0.070 eV, respectively[9]) but that increasingly larger amounts of covalent bonding and charge transfer are required to explain the bond in $ArAu^+$, $KrAu^+$, and $XeAu^+$ ($D_e = 0.464, 0.763$ and $1.25$ eV, respectively[9]).

However, experimental data are scarce. Kapur and Müller identified $NeAu^+$ in mass spectra of electric field-evaporated ions.[19] Schröder et al. detected $XeAu^+$ and $Xe_2Au^+$ in mass spectra by reacting $Au^+$ with $C_6F_6$ in the presence of xenon.[7] Neutral NeAu, ArAu, KrAu, and XeAu have been studied by resonance enhanced multiphoton absorption (see ref. 13 and references therein) and photodissociation.[20] Photoelectron spectra of $NgAu^-$ (Ng = Ne, Ar, Kr, Xe) in the gas phase have been reported by Gao et al..[21] Also worth mentioning are experiments in which vibrational spectra of small, neutral $Au_n$ or mixed $Au_nAg_m$ clusters complexed with a few krypton[22] or argon[23] atoms were recorded; the data indicated very strong interaction between gold clusters and noble gas atoms.[24-27]

In 2000 Seidel and Seppelt synthesized the first metal-xenon compound with direct gold-xenon bonds; the square planar structure of $[Xe_4Au]^{2+}$ was established by single-crystal diffraction.[28] Later the group reported compounds containing $[Xe_2Au]^{2+}$ and $[XeAu]^+$.[29-30] Stimulated by this work, Stace and coworkers succeeded in the mass spectrometric detection of $[Ar_nM]^{2+}$ (M = Cu, Ag, Au) cluster ions in the gas phase by passing a supersonic jet of argon clusters through a vapor of noble metals followed by electron ionization.[31] They observed $[Ar_nCu]^{2+}$ and $[Ar_nAg]^{2+}$ with $1 \leq n \leq 8$, and a much weaker signal of $[Ar_nAu]^{2+}$, $3 \leq n \leq 7$. Interesting features in the abundance distributions of $[Ar_nCu]^{2+}$ and $[Ar_nAg]^{2+}$ were maxima at $n = 4$ and sharp drops beyond $n = 6$, suggesting local anomalies in the stability of these ions. Singly charged $Ar_nAu^+$ ions were not mentioned in their report.[31]

Recently we reported mass spectra of helium nanodroplets doped with gold; cations containing up to 14 Au atoms and 85 of helium atoms were observed.[32] With the exception of this work, there are only two other published mass spectrometric studies of singly charged noble metal ions complexed with more than two noble gas atoms: One from our lab which suggest that $He_6Cu^+$, $He_{12}Cu^+$, and $He_{24}Cu^+$, are particularly stable.[33] And another one by Froudakis et al. who observed $Ne_nCu^+$ ($n \leq 24$) and $Ar_nCu^+$ ($n \leq 150$).[34] The abundance distributions indicated enhanced cluster ion stability at $n = 4$ and 12 for $Ne_nCu^+$ but at $n = 4$ and 6 for $Ar_nCu^+$. These observations were consistent with electronic structure calculations that revealed that for neon the first solvation shell has icosahedral symmetry while for Ar it has octahedral symmetry.[34] The differences were attributed to the different values of the size ratio

$$\sigma^* = R_{M-Ng}/R_{Ng-Ng} \qquad (1)$$

where $R_{M-Ng}$ is the equilibrium distance between the metal ion and the ligand while $R_{Ng-Ng}$ is the equilibrium distance between ligands. Within a hard-sphere model icosahedral packing is energetically preferred for $0.823 \leq \sigma^* \leq 0.951$ while octahedral packing is favored for $0.613 \leq \sigma^* \leq 0.707$.[35]

Here we report mass spectra of $Au^+$ complexed with up to 15 He, Ne, Ar, Kr, or Xe atoms, obtained by electron ionization of helium droplets doped with gold and noble gases. Local anomalies in the ion abundances hint at particularly stable cluster sizes.[36] Prominent anomalies in the abundance distributions of $Ar_nAu^+$, $Kr_nAu^+$, and $Xe_nAu^+$ are local maxima at $n = 2$. The high stability of these ions is expected; it derives from the covalent character of the bond.[1,5,10-11,37] Surprisingly, though, $Ne_2Au^+$ (but not $He_2Au^+$) forms a local abundance maximum as well. This seems to conflict with a theoretical study of $Ne_nAu^+$ at the CCSD(T) level which finds a higher dissociation energy for $Ne_3Au^+$ than for $Ne_2Au^+$,[12] and the general notion that the bonding in $NeAu^+$ and $Ne_2Au^+$ is entirely physical,[5,11-12] i.e. the attractive forces are due entirely to the ion-induced dipole term plus other induction and dispersion attractive terms.[11]

To the best of our knowledge, the current data present the first mass spectrometric study of solvation of a metal cation in all noble gases (ignoring radon). Thus, another aim is to establish trends in the structure and size of the solvation shell as the size ratio $\sigma^*$ decreases from He to Xe. Local maxima at $He_{12}Au^+$ and $Ne_{12}Au^+$ indicate icosahedral structure, consistent with their rather large estimated values of $\sigma^*$. $Ar_nAu^+$, on the other hand, seems to adopt octahedral structure, consistent with its much smaller $\sigma^*$ value. A computational study of $He_nAu^+$ and $Ar_nAu^+$, based on pairwise additive potentials computed by coupled-cluster calculations, corroborates this conclusion. Furthermore, the calculations and previously reported mass spectra[32] suggest that $He_nAu^+$ features two additional solvation shells of $I_h$ symmetry, similar to results obtained for other ions solvated in helium[38-39] or molecular hydrogen.[40-42]

## 2. Experiment

Helium nanodroplets were produced by expanding helium (Messer, purity 99.9999 %) through a 5 μm nozzle, cooled by a closed-cycle refrigerator, into vacuum. The stagnation pressure and nozzle temperature varied slightly from run to run; representative values were $P_0$ = 23 bar and $T_0$ = 9.6 K.[43] At these conditions the droplets contain an estimated average number of $10^5$ to $10^6$ helium atoms.[44] The expanding beam was skimmed by a 0.8 mm conical skimmer located 8 mm downstream from the nozzle and traversed an 8 cm long, differentially pumped pick-up cell into which Ne, Ar, Kr, or Xe (Messer, research grade, partial pressure about $1\times10^{-5}$ mbar) could be introduced. The droplet beam passed through another pick-up cell filled with gold vapor produced in a resistively heated oven. The temperature of the gold oven could not be measured, but it was varied in order to obtain the optimal conditions for formation of small $Ng_nAu^+$ cluster ions. $He_nAu^+$ ions are observed if no gas is introduced into the first pick-up cell.

The beam of doped helium droplets was collimated and crossed by an electron beam. The electron energy varied from run to run between 52 and 91 eV;[43] no major impact of the electron energy on the mass spectra is to be expected within this range. Cations were accelerated into the extraction region of a reflectron time-of-flight mass spectrometer (Tofwerk AG, model HTOF) with a mass resolution $m/\Delta m \approx 4500$ ($\Delta m$ = full-width-at-half-maximum). Further experimental details have been provided elsewhere.[45]

## 3. Theory

We have generated two-body potentials for the $Ng-Au^+$ and Ng-Ng interactions (Ng = He or Ar) based on the potential energy surfaces from *ab initio* calculations at the CCSD(T)/def2-TZVPP level. The potentials were corrected for basis set superposition error (BSSE) using the counterpoise scheme and the calculations were performed using the Gaussian 16 software.[46] Potential energy curves are displayed in Fig. S1. Classical molecular dynamics simulations of $Ng_nAu^+$ were run using the LAMMPS suite[47] and the potentials derived from the *ab initio* calculations were read as tabulated values.

The initial structures used in the simulations consisted of a randomized $Ng_{110}Au^+$ cluster with the Au ion located at the center. Simulations were carried out 1 K for He and 10 K for Ar for 50 ps (1 fs time step) to sample the potential energy surface of the systems and overcome barriers between different structures. At the end of each simulation, the structures were instantaneously optimized (at 0 K) to identify potential energy minima for each starting geometry. This process was repeated 100 times and the lowest potential energy structure was saved. A randomly chosen rare gas atom was then removed and the procedure was performed once again for the smaller cluster size. This was repeated for all smaller cluster sizes.

The structures of small clusters are analyzed visually while that of large clusters is analyzed by extracting their radial density distributions (rdf) in form of histograms that show the number of ligands as a function of their distance from $Au^+$. Time averages from the annealing processes are also shown in the supplementary material. Results will be presented in Section 5 and the Supplemental Information.

## 4. Results

Fig. 1 presents a mass spectrum of helium droplets doped with gold and argon. Gold is monoisotopic ($^{197}$Au, mass 196.967 u); argon is nearly monoisotopic ($^{40}$Ar, natural abundance 99.600 %, mass 39.962 u). The most prominent mass peaks in Fig. 1a, marked by triangles, are due to bare $Au_n^+$ (note the logarithmic ordinate). Secondary mass peaks are due to $Ar_nAu_m^+$; they are seen more clearly in Fig. 1b which employs a linear ordinate. Mass peaks due to $Ar_nAu^+$ are flagged by dots and connected by a solid line. Peaks due to $Ar_nAu_2^+$ and $Ar_nAu_3^+$ are flagged by diamonds and triangles and connected by dashed lines; the first three peaks in these series are off scale.

Extracting the abundance of $Ar_nAu^+$ ions from the mass spectrum poses no problems because of the absence of mass spectral interference with the well resolved $Ar_nAu_2^+$ and $Ar_nAu_3^+$ series and the very weak $He_n^+$ series. Analyzing mass spectra of the other noble gases is more challenging because they feature two (neon), five (krypton), or seven (xenon) naturally occurring isotopes above the 1 % level. A custom-designed software package is used to extract their ion abundance.[48] The software deconvolutes possible overlapping contributions to particular mass peaks by different ions and isotopologues. It automatically fits mass peaks, subtracts background signals, and explicitly considers isotopic patterns of all ions that are expected to contribute to a given peak.

The abundance distributions of $Ng_nAu^+$ ions are compiled in Fig. 2 for Ng = He through Xe (panels a through e, respectively). Significant anomalies are labeled; for most data points error bars are smaller than the symbols. All noble gases except helium feature a pronounced maximum at $n = 2$. $He_nAu^+$ and $Ne_nAu^+$ feature a pronounced maximum at $n = 12$. $Ar_nAu^+$ is the only system that shows significant anomalies at intermediate sizes, namely at $n = 6$ and 9.

We emphasize that these local anomalies are reproducible. For example, several abundance distributions of $He_nAu^+$ obtained under different experimental conditions (different helium droplet sizes and ion extraction conditions) have been presented in a previous report.[32] All distributions featured a distinct local maximum at $n = 12$.

It is worth mentioning that all Ng-Au cluster ions identified in our mass spectra are singly charged. Stace and coworkers observed $Ar_nAu^{2+}$, $3 \leq n \leq 7$ by passing a supersonic beam of argon clusters through gold vapor and subsequent electron ionization of the mixed clusters at 100 eV.[31] The researchers were solely interested in dications; they did not mention monocations in their report.

Results of our computational work will be presented and discussed in the next section; additional results will be provided as Supplemental Information.

## 5. Discussion
### 5.1 The significance of local anomalies in ion abundance distributions

The basic assumption is that, at experimental conditions prevalent in the present study, local anomalies in the abundance distribution of cluster ions $Ng_nAu^+$ correlate with corresponding features in their dissociation energy $D_n$ (i.e. the energy required to adiabatically remove one noble gas atom).[49-50] Cluster growth in a helium droplet by successive pick up of monomers is a statistical process which produces broad, featureless size distributions. However, subsequent electron ionization introduces a large amount of excess energy which results in fragmentation and possible enrichment of relatively stable cluster ions. For cations the process starts with the formation of $He^+$ in the droplet.[51-52] The positive charge may jump by resonant charge exchange to an adjacent helium atom. This hopping process is terminated either by the formation of $He_2^+$ or by charge transfer to the dopant. In the latter case, about 15 eV (the difference between the ionization energies of helium, 24.6 eV, and gold, 9.2 eV) will be released.[52-53] This energy exceeds the evaporation energy of bulk helium by a factor 24000.

The large excess energy leads to cluster ion ejection from the helium droplet and extensive fragmentation. Milliseconds elapse between ionization and mass analysis, providing ample time for the ions to cool by evaporation, enriching stable ions and depleting unstable ions. The relation between cluster abundance measured by mass spectrometry and (relative) dissociation energy is intricate.[54-55] The relation becomes semiquantitative for cluster ions whose heat capacity is much less than the classical value;[36] a comparison of measured ion abundances with calculated dissociation energies for $He_nC_{60}^+$ [56] and $He_nAr^+$ [38-39] supports that conclusion. The relation will be more complex for systems whose vibrational degrees of freedom are not frozen out as easily.[57] Qualitatively, however, it is clear that an abrupt decrease of the dissociation energy (i.e. $D_{n+1} \ll D_n$) will cause an enrichment of cluster ion $A_n^+$ at the expense of $A_{n+1}^+$. There are several different scenarios, such as a single cluster size that is particularly stable (a "magic" cluster) or particularly unstable with respect to its neighbors, or closure of a solvation shell where $D_n$ drops in a stepwise fashion. For simplicity we merely assume that a local maximum or abrupt drop in the ion abundance at a magic number $n$ indicates that the dissociation energy of $A_n^+$ is considerably larger than that of $A_{n+1}^+$.

Note that dissociation energies refer to zero temperature, i.e. they measure the energy difference between ground state structures. Evaporations, however, require finite temperatures, and one should really speak of free energies which may involve significant contributions from configurational entropies.[58] Entropic contributions are commonly neglected primarily because it is challenging to quantitatively account for them.

### 5.2 Magic numbers at $n = 2$

The abundance distributions of $Ne_nAu^+$, $Ar_nAu^+$, $Kr_nAu^+$, and $Xe_nAu^+$ feature a local maximum at $n = 2$, suggesting that the dissociation energy of $Ng_2Au^+$ greatly exceeds that of $Ng_3Au^+$. The abundance of $He_2Au^+$, on the other hand, is not enhanced. Are these results consistent with previous work?

Early research was focused on XeAu$^+$, the most stable of the diatomic ions. Pyykkö applied the CCSD(T) method and obtained a dissociation energy of $D_e$ = 0.91 eV.[1] Other values obtained with CCSD(T) and various basis sets are 1.31 eV,[7] 1.248 eV,[9] 1.33 eV,[10] 1.305 eV,[11] and 1.104 eV.[59] Electron correlation and relativistic effects have strong influence on the geometries and stabilities; they shorten the bond length and enhance the stability by some 50 %.[1,60]

Some studies have been devoted to diatomics containing noble gases lighter than Xe. Partly filled symbols in Fig. 3 represent $D_e$ values obtained by Yousef et al. with the RCCSD(T) procedure.[9] Several other theoretical groups have investigated these diatomics.[1,10-11,16-17] Their results are not included in Fig. 3 but they agree with Yousef's results within better than 10 %. The only exception are early results by Pyykkö[1] that were about 25 to 45 % below those of Yousef et al..[9]

A few theoretical studies have been devoted to triatomic Ng$_2$Au$^+$,[1,5] but for a discussion of the magic number at $n$ = 2 one needs dissociation energies that extend to $n$ = 3 or larger. The only such data have been calculated by Li et al. for Ne through Xe ($n \leq 3$) at the CCSD(T) level[12,37,59,61] and by Zhang et al. for Ar ($n \leq 6$) at the B3LYP level.[62] Their results are displayed in Fig. 3 by solid symbols connected by lines.

Fig. 3 reveals two striking features:
1. The stability of NgAu$^+$ and Ng$_2$Au$^+$ increases from He to Xe. The increase is less than a factor 2 from Ar to Kr to Xe, but much stronger (factor 7) from Ne to Ar.
2. For Ar, Kr and Xe the dissociation energy of Ng$_2$Au$^+$ is slightly larger than that of NgAu$^+$ but much larger (factor 2) than that of Ng$_3$Au$^+$. In contrast, the dissociation energy of Ne$_2$Au$^+$ is slightly weaker than that of its neighbors.[63]

Feature (1) reflects the notion that the bond in NgAu$^+$ and Ng$_2$Au$^+$ has significant covalent character for Xe and Kr, a slight covalent contribution for Ar,[1,5,10-11,37] but no covalent character for Ne.[5,11-12] However, as pointed out by Yousef et al.,[9] the large jump in the static electric dipole polarizability between Ne (0.396 Å$^3$) and Ar (1.64 Å$^3$) is the main reason for the strong increase in stability from NeAu$^+$ to ArAu$^+$.

Feature (2) poses a conundrum: Ne$_2$Au$^+$ forms a distinct local maximum in the ion abundance, similar to that of Ar$_2$Au$^+$ and the heavier noble gases, but its computed dissociation energy is not enhanced relative to that of Ne$_3$Au$^+$.[12] This suggests that theory underestimates the stability of Ne$_2$Au$^+$ or overestimates that of Ne$_3$Au$^+$.

It is tempting to conclude that the bond in NeAu$^+$ and Ne$_2$Au$^+$ is not entirely physical, contrary to conclusions drawn in previous studies.[5,11-12] A comparison with alkali ions solvated in noble gases, which are model systems for purely physical bonding, is instructive. For example, the structure and stability of alkali ions (Li$^+$, Na$^+$, K$^+$ and Cs$^+$) solvated in xenon has been explored by different theoretical approaches.[64-67] The calculated dissociation energies do not feature enhanced binding at $n$ = 2, even if pairwise additive potentials are augmented by three-body contributions that account for the interaction between the dipoles induced by the ion on the noble gas atoms.

### 5.3 The geometrical structure of He$_n$Au$^+$, Ne$_n$Au$^+$, and Ar$_n$Au$^+$

The abundance distributions in Fig. 2 feature magic numbers beyond $n$ = 2, most noticeably at $n$ = 12 for helium and neon, and $n$ = 6, 9 for argon. What information do these numbers convey? In general, magic numbers may be used to qualitatively test structural models if the proposed structure lends enhanced stability to particular sizes. For Au$^+$ solvated in He, Ne and Ar, the interaction is predominantly physical[5] and the interaction is reasonably well described by pairwise additive potentials. In this situation, the dissociation energy will feature an abrupt drop beyond the first coordination shell (closure of secondary shells, or partial closure of a shell, may give rise to additional anomalies). Hence, the magic number will count the number of ligands in the first solvation shell (from here on called $n_1$) and reflect, indirectly, its geometric structure.

In the simplest approximation, $n_1$ merely depends on the size of the ion relative to that of the ligand. It is convenient to characterize these two quantities by the size ratio $\sigma^* = R_{M-Ng}/R_{Ng-Ng}$ (see eq. 1). $\sigma^*$ slightly below 1 will favor icosahedral arrangements; structures with smaller coordination numbers such as a square antiprism (a twisted cube, symmetry $D_{4h}$), octahedron ($O_h$), or tetrahedron ($T_d$) will be preferred as the value of $\sigma^*$ decreases.[67] Prekas et al. have calculated, within a hard-sphere model, the values of $\sigma^*$ at which structural transitions occur.[35] Their results are summarized in Fig. 4 below the abscissa. For example, the solvated ion will fit tightly into an octahedral shell ($n_1$ = 6) if $\sigma^*$ = 0.707. The shell can no longer accommodate the ion if $\sigma^* > 0.707$, the ligands will no longer be in direct contact with each other, and the shell converts to a square antiprism ($n_1$ = 8) which offers a larger cavity.[68] The square antiprism can no longer accommodate the ion when $\sigma^*$ exceeds 0.823 and the ligands adopt icosahedral structure ($n_1$ = 12). The icosahedral shell will rupture if $\sigma^*$ exceeds 0.951 and a close-packed (crystalline face centered or hexagonal) structure will be favorable.

Prekas et al. have tested the predictions of the hard-sphere model with molecular dynamics simulations in which the ligand-ligand as well as the ion-ligand interactions were modeled by Lennard Jones potentials and the value of $\sigma^*$ was varied systematically.[35] The basic features of the hard-sphere model were confirmed but the structure of incomplete solvation shells was found to also depend on the cluster size.[35,69]

These trends have been confirmed in various experimental studies, for example for $Na^+$, $Al^+$, and $In^+$,[70] for $K^+$ and $Mg^+$,[35] $Li^+$,[71] and $Cu^+$ [34] solvated in various noble gases. Support also comes from theoretical studies employing more realistic two- or three-body potentials for the ion-ligand and ligand-ligand interactions; for example, for different alkali ions solvated in xenon[64-67].

We adopt these ideas in order to "predict" the structure and size of the first solvation shell in $Ng_nAu^+$. We equate $R_{M-Ng}$ with the equilibrium distance in $NgAu^+$ calculated by Yousef et al. for Ng = He,[9] and by Breckenridge et al. for Ne through Xe.[11] $R_{Ng-Ng}$ is equated with the nearest-neighbor distance in noble-gas crystals which adopt close-packed structures.[72] For Ne through Xe these crystal values deviate less than 3 % from the well-known equilibrium distance of the diatomics; for He this choice avoids bias arising from the exceedingly large zero-point motion in $He_2$.[73]

The values of $\sigma^*$ thus obtained are indicated in Fig. 4 above the abscissa. Based on this simple model one would expect an icosahedral (or close-packed) structure for the first solvation shell in $He_nAu^+$ and $Ne_nAu^+$, i.e. closure of the first solvation shell at $n_1 = 12$. For $Ar_nAu^+$ one would expect an octahedral solvation shell ($n_1 = 6$). These features agree with the anomalies observed in the ion abundance, Fig. 2.

Thus, a simple hard-sphere model[35] provides a rational for the anomalies observed in the abundance distributions of $Au^+$ solvated in He, Ne, and Ar. We discard the predictions of Fig. 4 for $Kr_nAu^+$ and $Xe_nAu^+$ because the presence of strong covalent bonding is inconsistent with the assumption of non-directional bonding.

For further insight we have performed molecular dynamics simulations of $He_nAu^+$ and $Ar_nAu^+$ assuming two-body ion-ligand and ligand-ligand interaction. The dissociation energies computed for small $Ar_nAu^+$ (not corrected for zero point energy) are displayed in Fig. 5b; results for $He_nAu^+$ are provided in the Supplemental Information, Fig. S4. Also included are results obtained by Zhang et al. at the B3LYP level for $n \leq 6$.[62] Our $Ar_nAu^+$ data indicate enhanced binding for $n = 6$ and 9, in agreement with the local anomalies in the abundance distributions (Fig. 5a). We emphasize, however, that our simulations do not account for covalent bonding; they cannot correctly predict the structure and energy of the ionic core; they cannot reproduce the anomaly at $n = 2$. $Ar_6Au^+$ has octahedral structure, in agreement with the B3LYP calculation by Zhang et al.. $Ar_7Au^+$ is a capped octahedron but $Ar_8Au^+$ converts to the square antiprism. The structures of the magic $Ar_9Au^+$ and $Ar_{10}Au^+$ vaguely resemble those of a capped and bicapped square antiprism, respectively. Geometries of select ions are displayed in the Supplemental Information (Fig. S5).

A convenient way to identify the presence and size of solvation shells is provided by radial density functions which are obtained by plotting the number of atoms versus their distance from the solvated ion. The histograms in Fig. 6 (red lines) represent rdfs of $He_{110}Au^+$ and $Ar_{108}Au^+$ at the end of a simulation, i.e. at 0 K (panels a and b, respectively). They reveal several distinct solvation shells. In order to determine the number of atoms in the shells we evaluate the cumulative number of atoms in the histograms *versus* distance, resulting in the blue curves.

For Ar (Fig. 6b), the total number of atoms within the first, first two, first three, and first four shells are 6, 10, 22 and 34, respectively. The number 10 does not quite agree with the anomaly at 9 that we identified in the plot of dissociation energies and ion abundances (Fig. 5). There are two possible reasons for this disagreement: i) The simulation of $Ar_{108}Au^+$ may have failed to produce the minimum energy structure. ii) The coordination number $n_1$ may slightly increase with cluster size because atoms in outer shells will exert additional forces on the atoms in the first shell.[74]

In contrast, the dissociation energy $D_n$ is determined from simulations of much smaller clusters ($D_9$, e.g., equals the difference between the total energies of Ar + $Ar_8Au^+$ and $Ar_9Au^+$). In the Supplement (Fig. S3) we show the rdf of $Ar_{108}Au^+$ averaged over the full (50 ps) duration of the simulation at 10 K. Peaks are broader, the third and higher peaks are no longer distinct, and the cumulative sum of atoms in the first two shells equals 9, consistent with the computed dissociation energies.

For He, the total number of atoms within the first, first two and first three shells are 12, 32, and 44, respectively. Abundance distributions of $He_nAu^+$ extending to $n = 85$ have been presented in a previous publication.[32] They do reveal, indeed, small but statistically significant abrupt drops on an otherwise smooth distribution at $n = 32$ and 44, in addition to the strong anomaly at 12. This same set of magic numbers (12, 32, 44) was more readily discernible in the ion abundance of $Ar^+$ solvated in He,[38] and $Cs^+$ or $H^-$ solvated in molecular hydrogen.[40-41] Theoretical studies of $He_nAr^+$ employing the Path Integral Monte Carlo method[39] and of $(H_2)_nH^-$ using CCSD(T)[42] reveal that these magic numbers reflect successive closure of three concentric solvation shells in which the ligands are placed at the vertices of platonic solids, namely an icosahedron, a dodecahedron, and an icosahedron.

Whereas the number of atoms in a solvation shell can be reliably surmised from the rdf, a structural characterization becomes challenging if the interaction between the ion and the ligands is weak and quantum effects dominate, as demonstrated in a recent study of $Cs^+$ solvated in helium.[75] In this system the Lindeman index which measures the root-mean-square bond length fluctuation indicates a fluid-like solvation shell. On the other hand, the first three helium layers around the smaller, more strongly bound $Ar^+$ appear to form a highly ordered solid.[39] The estimated ion-ligand distance of He-$Au^+$ ($R_{M-Ng}$ = 2.44 Å) is even shorter than that of He-$Ar^+$ (2.57 Å),[9] suggesting[75] that the interaction in $He_nAu^+$ is even stronger than in $He_nAr^+$, thus favoring solid-like order in $He_nAu^+$. It is also interesting to note that the first three peaks in the time-averaged rdf of $He_nAu^+$ (see Fig S2) remain distinct.

**Conclusion**

We have measured ion abundance distributions of $Au^+$ solvated in all noble gases, helium through xenon. Local maxima in the ion abundance for $Ar_2Au^+$, $Kr_2Au^+$, $Xe_2Au^+$ suggest that their dissociation energies are much larger than those of the corresponding $Ng_3Au^+$, in agreement with the covalent character of bonding in these systems.[1,5,10-11] and calculated dissociation energies.[12,37,59,61-62] Surprisingly, though, $Ne_2Au^+$ features an enhanced abundance as well, contrary to the notion that its bond is entirely physical and its dissociation energy is below that of $Ne_3Au^+$.[5,11-12] Anomalies in the ion abundance of $He_nAu^+$ and $Ne_nAu^+$ at $n$ = 12 and $Ar_nAu^+$ at 6 can be assigned to icosahedral and octahedral solvation shells, consistent with the relative size of the solvated ions. Model calculations with pairwise additive potentials provide further insight into the structure of these cluster ions. They suggest two additional solvation shells of icosahedral symmetry in $He_nAu^+$, consistent with weak but statistically significant features in the abundance distributions. Structural order in helium snowballs beyond the first solvation shell is the exception rather than the rule;[74] for a complete discussion one will have to consider not only the size ratio $\sigma^*$ and the strength of the ion-ligand interaction but the shape of the ion-ligand interaction potential as well.[39]

Supplemental Information
Shown are potential energy curves of the ion-ligand and ligand-ligand interaction, radial density functions, dissociation energies, and geometric structures of select sizes of $He_nAu^+$ and $Ar_nAu^+$.

Acknowledgement
This work was supported by the Austrian Science Fund, FWF (Projects I4130, P31149, and W1259), the Swedish Research Council (contract No. 2016-06625), and the European Commission (ELEvaTE H2020 Twinning Project, Project No. 692335).

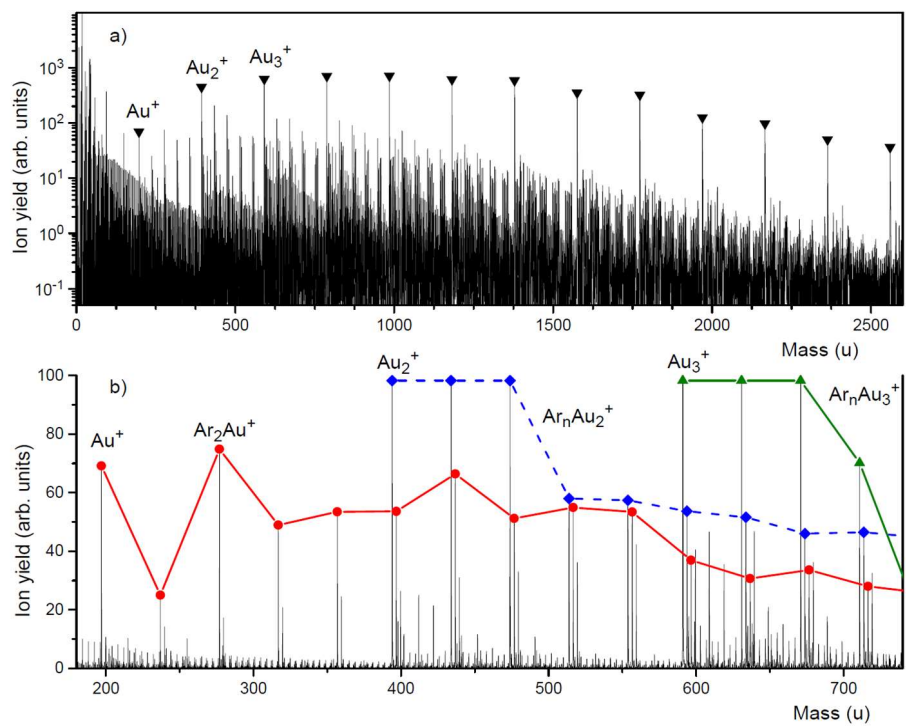

Fig. 1. Panel a: Mass spectrum of helium nanodroplets doped with gold and argon. Mass peaks due to bare $Au_m^+$, $1 \leq m \leq 13$, are marked. Panel b shows a section of the mass spectrum plotted with a linear ordinate. Ion series due to $Ar_nAu^+$, $Ar_nAu_2^+$, and $Ar_nAu_3^+$ are marked by dots, diamonds, and triangles, respectively.

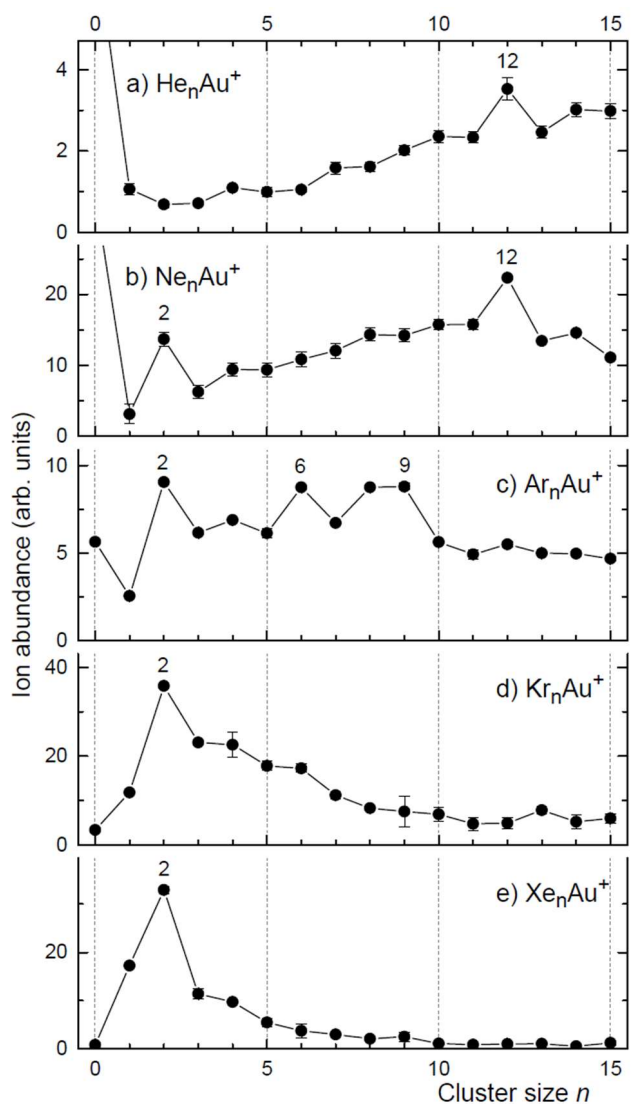

Fig. 2 Ion abundance of $Ng_nAu^+$ cluster ions (Ng = He, Ne, Ar, Kr, Xe). Significant local anomalies in the ion abundances are labeled.

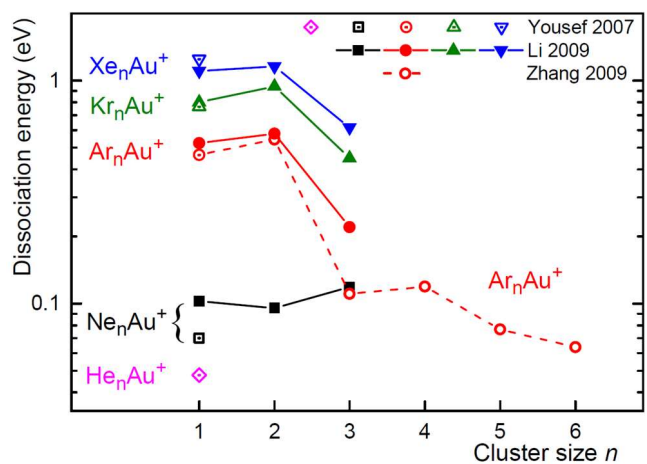

Fig. 3. Published dissociation energies calculated for $Ng_nAu^+$ ions. Partly filled symbols for diatomic $NgAu^+$ are from Yousef et al..[9] Solid symbols connected by solid lines for $Ng_nAu^+$ (n = 1, 2, 3; Ng = Ne, Ar, Kr, Xe) are from Li et al..[12,37,59,61] Open circles connected by a dashed line for $Ar_nAu^+$, $n \leq 6$, are from Zhang et al..[62]

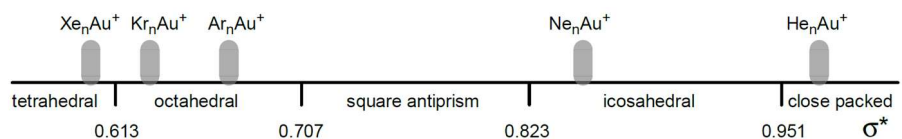

Fig. 4. Data below the abscissa indicate the structure of the first solvation shell for a metal ion (M$^+$) solvated in a noble gas (Ng), predicted within a hard-sphere model as a function of $\sigma^* = R_{\text{M-Ng}}/R_{\text{Ng-Ng}}$ where $R_{\text{M-Ng}}$ is the distance between the ion and the ligand, and $R_{\text{Ng-Ng}}$ is the distance between adjacent ligands.[35] $\sigma^*$ values for M$^+$ = Au$^+$ and Ng = He, Ne, Ar, Kr, Xe are indicated above the abscissa; they are estimated from literature data (see text for details).

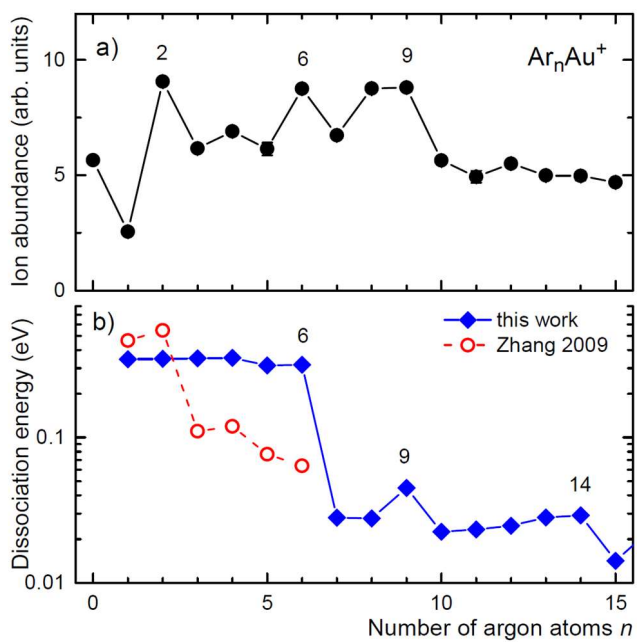

Fig. 5. Comparison of the ion abundance of Ar$_n$Au$^+$ (panel a) with calculated dissociation energies (panel b). Diamonds (this work) indicate data obtained using pairwise additive potentials. Values represented by open circles were computed at the B3LYP theoretical level by Zhang et al..[62]

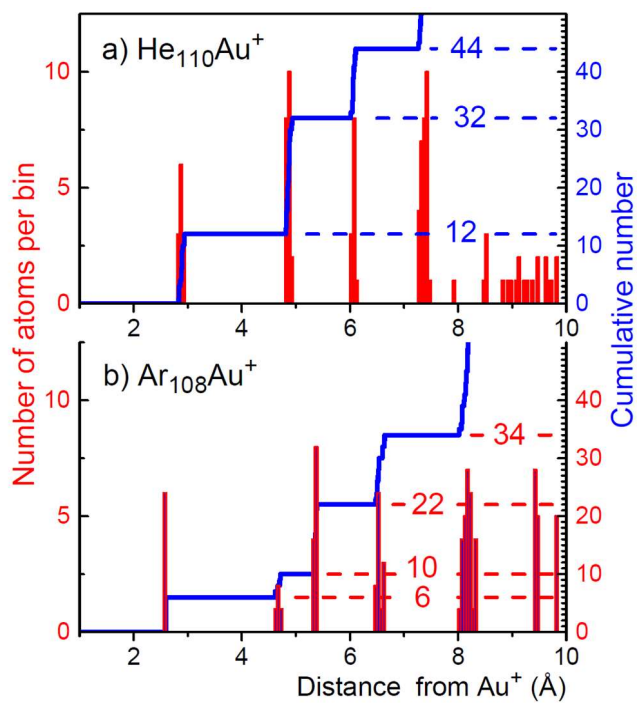

Fig. 6. The histograms (red lines) represent the computed radial helium density around Au$^+$ in He$_{110}$Au$^+$ and Ar$_{108}$Au$^+$ (panels and b, respectively; values are specified on the left ordinate). Blue lines represent integrated histograms, i.e. the number of He atoms within a sphere of radius $r$ around Au$^+$ (right ordinate). Values in the graph specify the cumulative number of atoms within the first, second, third,… solvation shell.